\documentclass[aps,prl,twocolumn,longbibliography]{revtex4-2}
\usepackage{hyperref}
\hypersetup{
    colorlinks = true,
    linkcolor = [rgb]{0.70, 0.13, 0.13},
    citecolor = [rgb]{0.25, 0.41, 0.88},
    urlcolor = [rgb]{0.25, 0.41, 0.88}
}
\usepackage{amsmath}
\usepackage{amsfonts}
\usepackage{amssymb}
\usepackage{pifont}
\usepackage{mathtools}
\usepackage{bm}
\usepackage{cases}
\usepackage{braket}
\usepackage[version=4]{mhchem}
\usepackage{graphicx}
\allowdisplaybreaks%

\makeatletter
\newcommand*{\sumcirclearrowleft}{%
  \DOTSB
  \mathop{
    \mathchoice
      {\rlap{\kern.25em\rotatebox[origin=c]{-90}{$\circlearrowleft$}}{\sum}}
      {\vcenter{\rlap{\kern.2em\rotatebox[origin=c]{-90}{$\scriptscriptstyle\circlearrowleft$}}}{\sum}}
      {\sum}{\sum}
  }\slimits@
}

\newcommand*{\sumcirclearrowright}{%
  \DOTSB
  \mathop{
    \mathchoice
      {\rlap{\kern.25em\rotatebox[origin=c]{90}{$\circlearrowright$}}{\sum}}
      {\vcenter{\rlap{\kern.2em\rotatebox[origin=c]{90}{$\scriptscriptstyle\circlearrowright$}}}{\sum}}
      {\sum}{\sum}
  }\slimits@
}
\makeatother

\begin{document}

\title{Emergent Berezinskii-Kosterlitz-Thouless and Kugel-Khomskii physics \\ in the triangular lattice bilayer colbaltate}

 \author{Gang V. Chen}
\affiliation{International Center for Quantum Materials, School of Physics, Peking University, Beijing 100871, China}
\affiliation{Collaborative Innovation Center of Quantum Matter, 100871, Beijing, China}
 
\begin{abstract}
Motivated by the experiments on the triangular lattice bilayer colbaltate K$_2$Co$_2$(SeO$_3$)$_3$, 
we consider an extended XXZ model to explore the underlying physics from a couple observation. 
The model is composed of interacting Co$^{2+}$ dimers on the triangular lattice, 
where the Co$^{2+}$ ion provides an effective spin-1/2 local moment via the spin-orbit coupling 
and the crystal field effect. The intra-dimer interaction is dominant and would simply 
favor the local spin singlet, and the inter-dimer interactions compete with the inter-dimer interaction, 
leading to rich behaviors. With the easy-axis anisotropy, it is shown that, in the ground state manifold 
of the intra-dimer Ising interaction, the system realizes an effective quantum Ising model, 
where the ground state is either a 3-sublattice order with a mixture of antiferromagnetic 
Ising order and a valence-bond spin singlet or Ising disordered. 
The finite temperature regime realizes the Berezinskii-Kosterlitz-Thouless physics. 
To explore the full excitations, we incorporate the excited state manifold of the intra-dimer 
Ising interaction and establish the emergent Kugel-Khomskii physics. Thus, 
the triangular lattice bilayer colbaltate is an excellent platform to explore 
the interplay between geometrical frustration and anisotropic interactions
as well as the emergent effective models and the resulting physics.
\end{abstract}

\maketitle
 
Frustrated magnets have been an active field in modern condensed matter physics~\cite{Frustrated}.  
Frustration arises from competing interaction that cannot be optimized simultaneously.
A common source of frustration is the geometrical frustration on frustrated lattices
such as triangular, kagom\'{e}, and pyrochlore lattices. The study of frustrated magnets 
on these frustrated lattices has contributed to a large portion of topics. 
Theoretically, frustration generates a large number of degenerate and nearly degenerate 
many-body states on which the sub-dominant interactions are reduced to effective interaction 
and then lead to rather interesting and exotic behaviors. Thus, the ingredient of frustration 
has inspired many contrived theoretical models and works in last two decades. 
On the other hand, the abundance of frustrated magnetic materials has provided 
the experimental motivation for new frustrated models and phenomena out of the frustration. 

One ongoing interest on the frustrated materials is about the spin-orbit-coupled Mott insulators~\cite{Witczak_Krempa_2014}. 
These materials are often the $4d/5d$ transition metal compounds and $4f$ rare-earth magnets. 
In these systems, due to the spin-orbit coupling and the electron correlation, the magnetic interactions
are often quite anisotropic, both in the spin space and the position space. The resulting models may 
often differ significantly from simple Heisenberg models, and these anisotropic models may carry
strong frustration even on unfrustrated lattices. The representative example is the honeycomb lattice
Kitaev model that has the relevance on various iridates and $\alpha$-RuCl$_3$~\cite{Takagi_2019}. 
With both anisotropy and frustration, it is natural to expect that these anisotropic and frustrated 
models on geometrically frustrated lattices could bring more unexpected behaviors. 
Much progress has been made in the rare-earth pyrochlore magnets~\cite{Rau_2019} 
and the $4d/5d$ Kitaev materials~\cite{Takagi_2019}. 

% summarize Co-magnets

\begin{figure}[b]
\includegraphics[width=7.5cm]{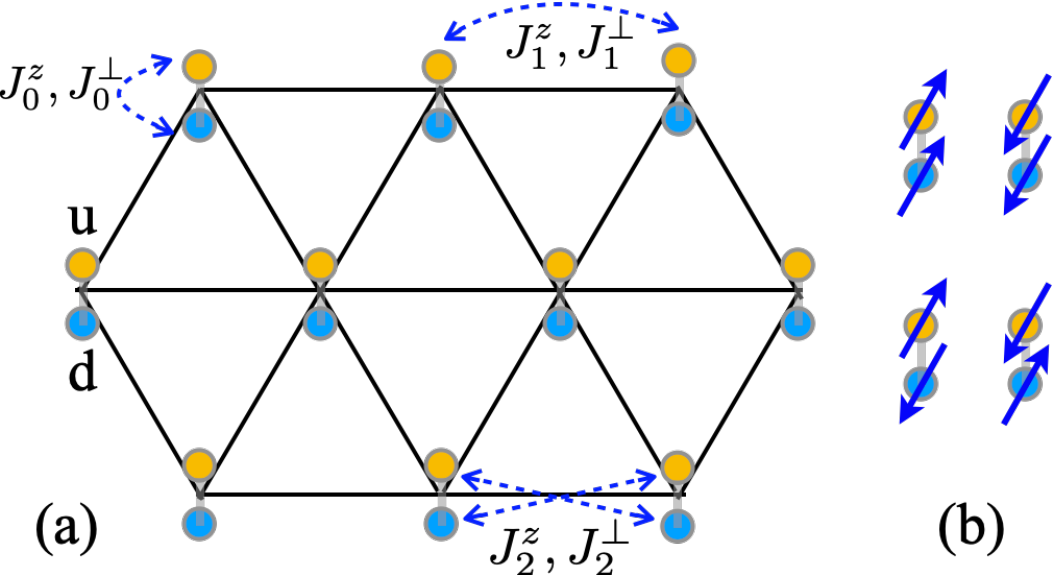}
\caption{ (a) The Co$^{2+}$ dimers form a triangular lattice. 
(b) Two anticollinear and two collinear spin configurations on the dimer, where the spins point along
 the $\pm \hat{z}$ direction.}
\label{fig1}
\end{figure}

Some of the recent attention has been directed to the colbaltates. One motivation was 
to explore the Kitaev interaction and Kitaev spin liquid among the honeycomb colbaltates
where the spin-orbit coupling plays a similar role as the one for the iridates and induces the 
anisotropic interactions, and the progress has been made in the honeycomb magnets Na$_2$Co$_2$TeO$_6$ 
and Na$_3$Co$_2$SbO$_6$~\cite{PhysRevB.97.014407,PhysRevB.97.014408}. 
In fact, the spin-orbit coupling can sometimes play an important role for the underlying physics 
even in the $3d$ transition metal compounds, and this differs from elements to elements and
depends on the crystal environments.  
Moreover, several colbaltates are long known to be good realizations of the anisotropic models 
such as the quantum Ising model as the extreme. 
These include the quasi-1D Ising magnets CoNb$_2$O$_6$~\cite{Coldea_2010,PhysRevX.4.031008,Morris_2021}, 
BaCo$_2$V$_2$O$_8$~\cite{PhysRevLett.120.207205,Suga_2008,Wang_2018},
and SrCo$_2$V$_2$O$_8$~\cite{PhysRevLett.123.067203} where not only the external field 
gives a useful tunability of the quantum phases but also the system could produce emergent 
and exotic excitations associated with the criticality~\cite{Coldea_2010,PhysRevLett.127.077201}.

We are inspired by the experiments 
on the Co-dimer triangular lattice antiferromagnet K$_2$Co$_2$(SeO$_3$)$_3$~\cite{PhysRevB.102.224430}. 
As it is shown in Fig.~\ref{fig1}, the two neighboring Co$^{2+}$ ions form a dimer along $z$ direction,
and these dimers are then organized into a triangular lattice. It can be equivalently viewed as 
a bilayer triangular lattice, and the interlayer separation is quite short compared to the lattice constant. 
The magnetic susceptbility and the magnetization measurements clearly indicate the easy-axis spin anisotropy,
and the magnetic entropy is consistent with the effective spin-1/2 degree of freedom for each Co$^{2+}$ ion. 
The presence of the 1/3 magnetization plateau is compatible with the predominant Ising interaction or Ising-like 
moments. Given the previous success of the triangular lattice XXZ model on the spin supersolid compound
Na$_2$BaCo(PO$_4$)$_2$~\cite{pnas.2211193119,magnetocaloric,jia2023quantum} 
and the 120-degree ordered compound Ba$_3$CoSb$_2$O$_9$~\cite{PhysRevLett.116.087201},    
we consider an extended XXZ model with the easy-axis anisotropy on this bilayer triangular lattice.  
In the absence of the magnetic field, the intra-dimer Ising interaction favors 
the anticollinear spin configuration. Within this manifold, we show that the remaining 
interactions effectively realize a quantum Ising model on the triangular lattice and then gives rise to the
quantum order by disorder effect and the Berezinskii-Kosterlitz-Thouless (BKT) physics at
finite temperatures. Although the original XXZ model has a U(1) symmetry, 
the emergent BKT physics has no connection with the U(1) spin symmetry of the spin model. 
For the magnetic excitations and the magnetization plateau states,
  the states out of the intra-dimer anticollinear spin states are naturally involved. 
After incorporating these ``excited'' collinear states in Fig.~\ref{fig1}(b), we end up 
with the Kugel-Khomskii-like model and thus deal with the Kugel-Khomskii (KK) physics. 

\begin{table}[t]
\begin{tabular}{cccccccc}
\hline\hline
$S^z_{ui}$ & $S^x_{ui}$ & $S^y_{ui}$ & $S^z_{di}$ & $S^x_{di}$ & $S^y_{di}$ & $ S^z_{ui} S^z_{di}$ & $ S^x_{ui} S^x_{di} +S^y_{ui} S^y_{di}  $ 
\vspace{2mm}
 \\
$\tau^z_i$ & 0 & 0 & $-\tau^z_i$ & 0 & 0 &$-\frac{1}{4}$ & $\tau^x_i  $
\vspace{2mm}
\\
$\tau^z_i$ & $2 \tau^x_i \mu_i^x $ & $2 \tau^y_i \mu^x_i$ & $2\tau^z_i \mu^z_i$ & $ \mu^x_i$ & $2 \tau^z_i \mu^y_i$ & $\frac{\mu^z_i}{2}$ &  $\tau^x_i  (\frac{1}{2} - \mu^z_i)$
\vspace{2mm}
\\
\hline\hline
\end{tabular}
\caption{The list of spin operators on each dimer under different representations. 
The first line is the original spin operator on the local lattice site. 
The second line is in terms of the pseudospin $\tau_i^{\mu}$ after being projected onto the ground state manifold 
of the intra-dimer Ising interaction. 
The third line is in terms of the pseudospin $\tau_i^{\mu}$ and $\mu_i^{\nu}$ after including all the local states of the dimer.
The prescription is described in the main text. 
}
\label{tab1}
\end{table}

To explore the interplay between geometrical frustration and anisotropic interaction, 
we first write down the extended XXZ model with the spin-1/2
moments on the dimer triangular lattice,
\begin{eqnarray}
H &=& H_0 + H_{u} + H_{d} + H_{ud} + H_{\text{Z}}, 
\label{eq1}
\end{eqnarray}
where the five different terms are given as 
\begin{eqnarray}
H_0 &=& \sum_i \big[ J_{0}^z S^z_{ui} S^z_{di} + J_0^{\perp} ( S^x_{ui} S^x_{di} +S^y_{ui} S^y_{di} ) \big]  ,
\\
H_{u} &=& \sum_{\langle ij \rangle }  \big[  J_{1}^z S^z_{ui} S^z_{u j} + J_{1}^{\perp} ( S^x_{ui} S^x_{uj} +S^y_{ui} S^y_{uj} )  \big] ,
\\
H_{d} &=& \sum_{\langle ij \rangle }  \big[  J_{1}^z S^z_{di} S^z_{d j} + J_{1}^{\perp} ( S^x_{di} S^x_{dj} +S^y_{di} S^y_{dj} )  \big] ,
\\
H_{ud} &=& \sum_{\langle ij \rangle }  \big[  J_{2}^{z} S^z_{ui} S^z_{d j} + J_{2}^{\perp} ( S^x_{ui} S^x_{dj} +S^y_{ui} S^y_{dj} )  \big] 
\nonumber \\
&& \quad + \big[  J_{2}^{z} S^z_{di} S^z_{uj} + J_{2}^{\perp} ( S^x_{di} S^x_{uj} +S^y_{di} S^y_{uj} )  \big] ,
\\
H_{\text{Z}} & = & \sum_i -B ( S^z_{ui} + S^z_{di} ) .
\end{eqnarray}
Here $H_0$ is the intra-dimer interaction, $H_u$ is the inter-dimer interaction within the up layer,
$H_d$ is the inter-dimer interaction within the down layer, 
$H_{ud}$ is the inter-dimer interaction between the up and down layers,
and $H_{\text{Z}}$ is the Zeeman coupling for the field along the $z$ direction. 
The spin operator $S^{\mu}_{ui}$ ($S^{\mu}_{di}$) refers to the spin component $\mu$ 
at the site $i$ from the up (down) layer of the dimer triangular lattice. 
The spin interactions are labeled in Fig.~\ref{fig1}(a).

From the lattice geometry, the intra-dimer interaction should be dominant.
From the easy-axis anisotropy, the intra-dimer Ising interaction $J^z_0$ is 
stronger than the $xy$ exchange $J^{\perp}_0$. Thus, throughout this work, 
we will treat the intra-dimer Ising interaction as the dominant interaction. 
For the inter-dimer interaction, we still consider the easy-axis anisotropy. 
Moreover, we expect $J_1^z \gg J_2^z$ from the different exchange paths.

\begin{figure}[b]
\includegraphics[width=8.5cm]{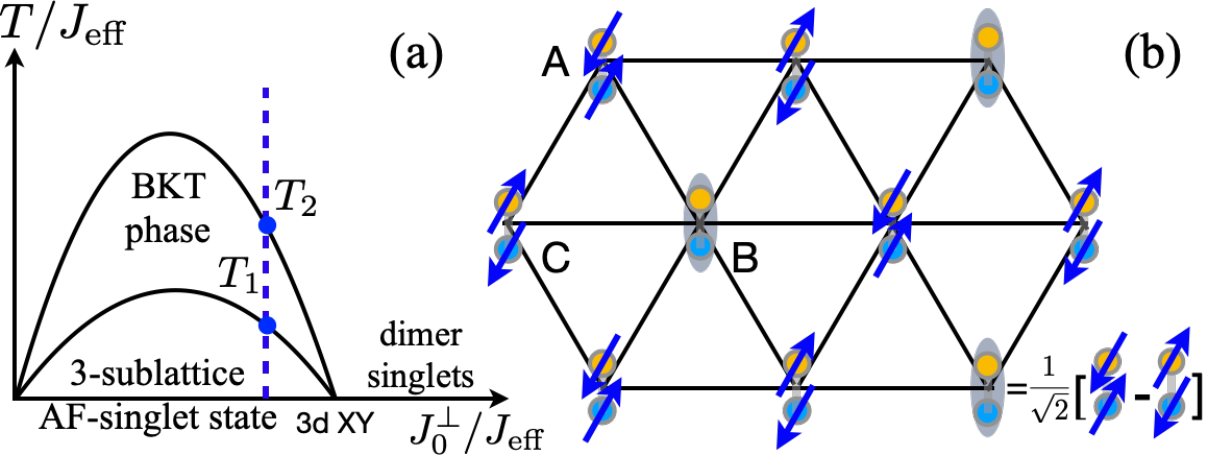}
\caption{(a) The phase diagram of quantum Ising model 
in Eq.~\eqref{eqHeff} (adapted from Ref.~\onlinecite{PhysRevB.63.224401}), 
where ${J_{\text{eff}} = 2(J_1^z- J_2^z)}$. 
(b) The 3-sublattice AF-singlet state in the limit of weak transverse field for the dimer triangular lattice. 
}
\label{fig2}
\end{figure}

\begin{figure*}
\includegraphics[width=4cm]{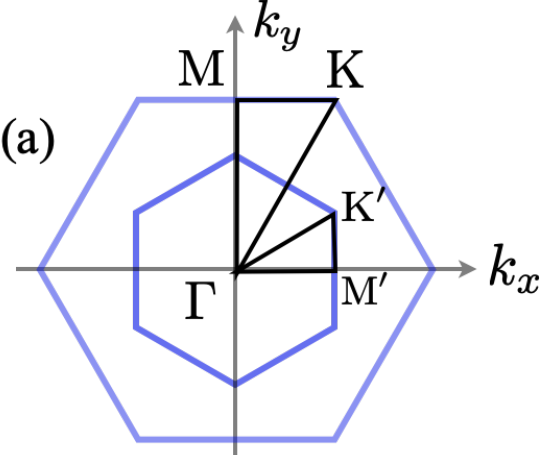}
\includegraphics[width=4.5cm]{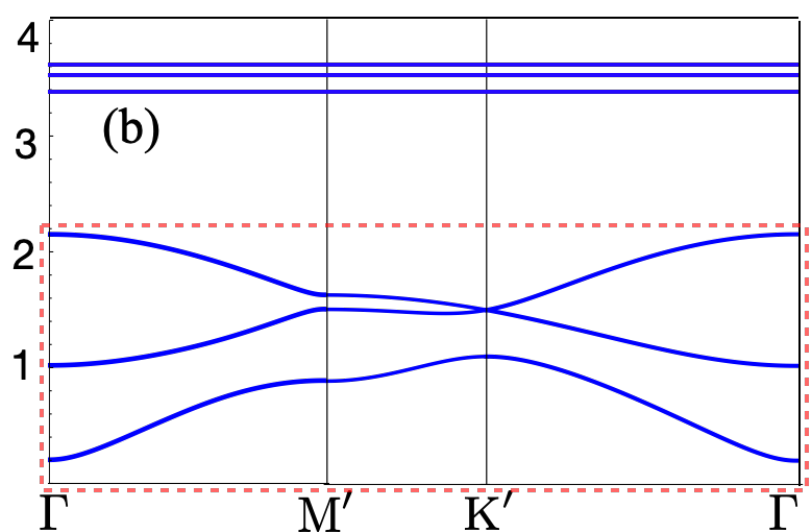}
\includegraphics[width=4.5cm]{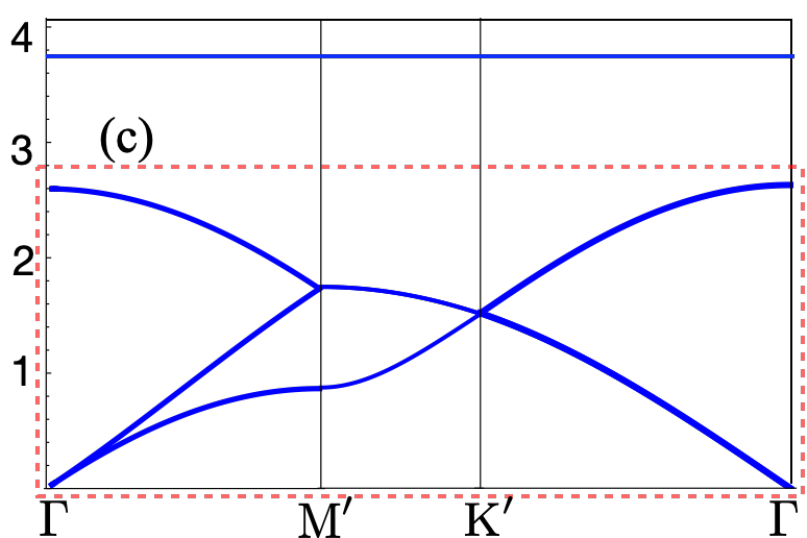}
\includegraphics[width=4.5cm]{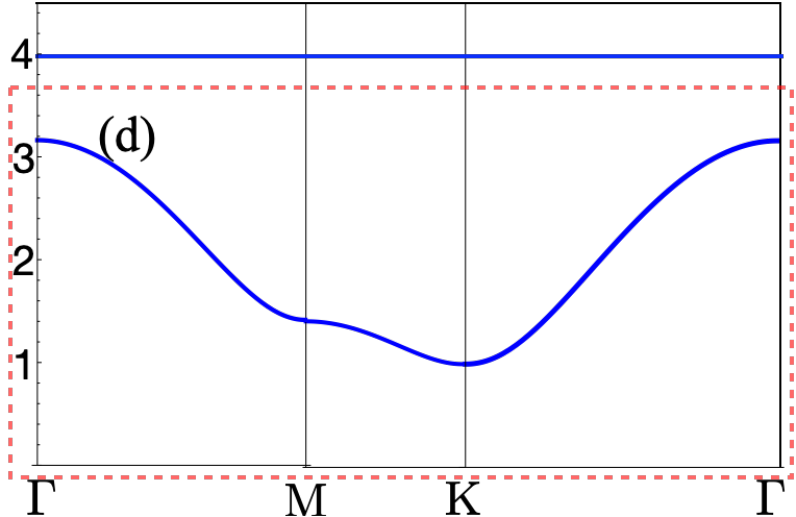}
\caption{The full elementary excitations in different zero-temperature phases for the minimal model $H_{\text{min}}$. 
(a) The lattice Brillouin zone and the magnetic Brillouin zone with the high symmetry momentum lines.
(b) The elementary excitations in the three sublattice ordered state, where we choose ${J_{\text{eff}}=1, J_0^{\perp} = 1.1, J_0^z= 6}$. 
(c) The elementary excitations at the criticality, where we choose ${J_{\text{eff}}=1, J_0^{\perp} = 1.5, J_0^z= 6}$.
(d) The elementary excitations for the dimer singlet, where we choose ${J_{\text{eff}}=1, J_0^{\perp} = 2, J_0^z= 6}$.
The spectra in the (red) dashed box are the elementary excitations for the effective model $H_{\text{eff}}$ (within the ${\mu^z=-1/2}$ manifold), 
and the rest are excitations out of the ${\mu^z=-1/2}$ manifold.  
The flat excitation would become weakly dispersive once the inter-dimer transverse coupling is included,
and the bandwidth of the $a$-bosons (i.e. excitations within the ${\mu^z=-1/2}$ manifold) should be suppressed by the renormalized
order parameter. 
}
\label{fig3}
\end{figure*}

We consider the zero-field case. The ground states of the intra-dimer Ising interaction 
are doubly degenerate for each dimer, and we introduce an effective pseudospin-1/2 states 
to label them as
\begin{eqnarray}
| \tau^z_i =+ \frac{1}{2}\rangle & = & |{S^z_{ui} = + \frac{1}{2}, S^z_{di} = -\frac{1}{2} } \rangle ,
\label{eq7}
\\
| \tau^z_i =- \frac{1}{2}\rangle & = & |{S^z_{ui} = - \frac{1}{2}, S^z_{di} = + \frac{1}{2} } \rangle .
\label{eq8}
\end{eqnarray}
We define a projection operator to this reduced local Hilbert space, 
$P= \prod_i \sum_{\alpha}|\tau^z_i =\alpha \rangle \langle \tau^z_i = \alpha |$.
The effective model in this reduced Hilbert space is $H_{\text{eff}} = P H P$. 
Different spin operators in this reduced Hilbert space is expressed in Tab.~\ref{tab1}. 
The effective model is given as a quantum Ising model with
\begin{eqnarray}
H_{\text{eff}} =\sum_i J_{0}^{\perp} \tau^x_i + \sum_{\langle ij \rangle} 
2(J_{1}^{z} -J_{2}^{z} ) \tau^z_i \tau^z_j - \frac{NJ_{0}^z}{4}  ,
\label{eqHeff}
\end{eqnarray}
where $N$ is the total number of dimers. This is a good approximation as long as $J_0^z$ is large enough. 
Remarkably, the inter-dimer transverse exchanges all disappear in this projection as they
do not connect the singlet manifolds of neighboring dimers. 
Unlike CoNb$_2$O$_6$~\cite{Coldea_2010,PhysRevX.4.031008,Morris_2021} and
BaCo$_2$V$_2$O$_8$~\cite{PhysRevLett.120.207205,Suga_2008,Wang_2018}
whose transverse field is applied externally, the transverse field here 
is of the intrinsic origin~\cite{PhysRevResearch.1.033141}. 
From $H_{\text{eff}}$, one can obtain many properties of the system. 
The phase diagram of this model is well-known from Ref.~\onlinecite{PhysRevB.63.224401} 
and is adapted to our problem in Fig.~\ref{fig2}(a), 
where we set ${J_{\text{eff}} = 2(J_{1}^{z} -J_{2}^{z})}$. 
There are two phases at zero temperature. With the large transverse field, 
the ground state is polarized 
in the negative $\tau^x$ direction, which is equivalent to the local singlet for each dimer.
When the effective exchange becomes large, a 3-sublattice order is favored instead, 
leading to the Bragg peaks at the ${\bf K}$ points of the Brillouin zone. 
This 3-sublattice order is better visualized in terms of the original spin configuration 
instead of the pseudospin $\tau_i$. We depict this 3-sublattice spin configuration
in the limit of weak transverse field in Fig.~\ref{fig2}(b). 
Since $\tau^x$ is always polarized 
by the transverse field, each dimer is generically a mixture of the antiferromagnetic Ising 
configuration and the valence bond spin singlet. This state is then refereed as 3-sublattice AF-singlet state,
and was not previously known.
The property in this 3-sublattice state can be captured by 
a Weiss mean-field theory (MFT) with
\begin{eqnarray}
H_{\text{eff}} \rightarrow H_{\text{MF}} = \sum_i J_{0}^{\perp} \tau^x_i + \sum_{\langle ij \rangle} J_{\text{eff}}  \,
\tau^z_i  \langle  \tau^z_j  \rangle  ,
\end{eqnarray}
where $ \langle  \tau^z_j  \rangle $ is the mean-field order parameter. 
The details are discussed in the Supplementary materials~\cite{Supple}. 
The excitations are plotted in Fig.~\ref{fig3}, and there exists 
a mini-gap in the spectrum. This is known as the pseudo-Goldstone mode that results from the 
quantum order by disorder effect. The effective model does not have the continuous symmetry
and the magnetic order does not break the U(1) symmetry of the original spin model in Eq.~\eqref{eq1}, 
so the excitation spectrum is fully gapped. The mini-gap indicates a weak stability of the ground state
against the magnetic field, and thus suggests a tiny magnetization plateau in the weak field limit. 

The quantum correction or renormalization of the order parameter is not considered in this MFT 
and thus the bandwidth of the excitations is enhanced compared to the reality. 
The overall dispersion does not significantly depend on the order parameter. 
The critical value of $J_{0}^{\perp}/J_{\text{eff}}$ in the mean-field theory is 1.5
while the numerical result from quantum Monte Carlo is ${\sim 0.82}$~\cite{PhysRevB.68.104409,PhysRevResearch.2.043013}.  
It is important to mention that the excitations are expressed in the variable of 
${\boldsymbol{\tau}}_i$.  The $\tau^z$-$\tau^z$ correlation should create the magnon excitations as well
as two-magnon continuum in the 3-sublattice AF-singlet state, and creates the 
magnon-like dispersive excitation in the Ising disordered state. The local probe such as NMR and 
$\mu$SR should also be useful to detect the dynamics in this reduced Hilbert space. 
The $\tau^x$ component is the bond operator in terms of the original spin variables 
and is even under time reversal. A useful probe is likely to be the X-ray pair distribution 
function measurement. 

At finite temperatures above the 3-sublattice AF-singlet phase, the order melts via the two-step process
before entering the high-temperature paramagnet (see Fig.~\ref{fig2}(a))~\cite{PhysRevB.16.1217,PhysRevB.63.224401}. 
The two finite-temperature phase transitions are 
of the BKT type, and the precise locations of the transitions, $T_1$ and $T_2$ (especially the higher temperature one), 
might be a bit difficult to detect, and show up as broad peaks in specific heat. 
Below $T_1$, the magnetic order emerges 
as a rigid order with a clear gap. 
Between $T_1$ and $T_2$, the system is in a BKT phase with an emergent U(1) symmetry
and the spin correlations are of the power law form. It is an algebraically ordered regime
but has no true long-range order with~\cite{PhysRevLett.115.127204} 
\begin{eqnarray}
\langle \tau^z_i \tau^z_j \rangle \sim \frac{ \cos [{\bf Q} \cdot 
( {\bf r}_i - {\bf r}_j) ] }{   | {\bf r}_i - {\bf r}_j |^{\eta(T)} },
\end{eqnarray}
where the wavevector ${\bf Q}$ corresponds to the ${\bf K}$ point, and
${\eta \in (1/9,1/4)}$ is temperature dependent. It was even predicted that, the uniform susceptibility 
in $\tau^z$ is singular in the power-law in the zero-field limit if a uniform field is applied to the pseudospin 
$\tau^z$ components~\cite{PhysRevB.97.085114}. This corresponds to a weak $\Gamma$ point peak in 
the $\tau^z$-$\tau^z$ correlation, and it is not true ferromagnetic order in $\tau^z$. 
If the system is close to the quantum phase transition between the 3-sublattice AF-singlet state and the dimer singlets, 
the system would be controlled by the critical mode at the finite temperatures,
and have a $T^2$ specific heat.

To obtain the full excitations, we ought to return to the original spin model in Eq.~\eqref{eq1}. 
Without losing connection to the effective model in Eq.~\eqref{eqHeff}, we incorporate the 
local excited states in Fig.~\ref{fig1}(b) as 
\begin{eqnarray}
| \tau^z_i = +\frac{1}{2} , \mu^z_i =+ \frac{1}{2} \rangle & = & | S^z_{ui} = +\frac{1}{2}, S^z_{di} = +\frac{1}{2} \rangle ,
\\
| \tau^z_i =- \frac{1}{2} , \mu^z_i =+ \frac{1}{2} \rangle & = & | S^z_{ui} =- \frac{1}{2}, S^z_{di} =- \frac{1}{2} \rangle ,
\end{eqnarray}
and the ground states in Eq.~\eqref{eq7} and Eq.~\eqref{eq8} are then treated as 
$|{ \mu^z_i =- {1}/{2} }\rangle$ with the same $\tau^z$'s. In this representation of the local Hilbert space, 
the spin variables are re-expressed in terms of the pseudospins $\tau^{\mu}$ and $\mu^{\nu}$ and are 
listed in Tab.~\ref{tab1}. The original model in Eq.~\eqref{eq1} now becomes 
\begin{widetext}
\begin{eqnarray}
H &=& \sum_i \big[ \frac{1}{2} J_{0}^{z} \mu^z_i + J_{0}^{\perp} \tau^x_i (\frac{1}{2} - \mu^z_i) \big] 
  + \sum_{\langle ij \rangle} 
\Big[  
  4J_{1}^z  \tau^z_i \tau^z_j (\frac{1}{4} +  \mu_i^z \mu_j^z) + 4J_1^{\perp} \big[ 
( \frac{1}{4}+ \tau^x_i \tau^x_j ) \mu_i^x \mu_j^x + \tau^y_i \tau^y_j \mu_i^x \mu_j^x +    \tau^z_i \tau^z_j \mu_i^y \mu_j^y 
  \big]
  \Big]
  \nonumber \\
&&  + \sum_{\langle ij \rangle} 
\Big[ 2 J_{2}^{z}  \tau^z_i \tau^z_j (\mu^z_i + \mu^z_j) 
 + 4J_{2}^{\perp} \big[\frac{1}{2} (\tau^x_i + \tau^x_j) \mu^x_i \mu^x_j + \tau^y_i \tau^z_j \mu^x_i \mu^y_j 
 + \tau^z_i \tau^y_j \mu^y_i \mu^x_j  \big] \Big]
 -\sum_i B\, \tau_i^z (1 + 2 \mu_i^z) .
\end{eqnarray}
It is in the form of the Kugel-Khomskii model except that the local degrees of freedom are from the dimer instead 
of the spin and orbitals~\cite{PhysRevLett.81.2356}. 
Thus, the system belongs to the so-called ``multiflavor Mott insulator''~\cite{Chen2021MultiflavorMI}. 
Based on this model, one can proceed and calculate the full excitations of the 3-sublattice AF-singlet state 
as well as the dimer singlet state. The presence of $J_1^{\perp},  J_2^{\perp}$ 
couplings may generate more phases if they become comparable to the Ising exchanges. 
To avoid further complication, we set ${J_1^{\perp} = J_2^{\perp} =0}$ as 
they might be weaker compared to their corresponding Ising couplings. 
The simplified model is the minimal model for our calculation below and is given as
\begin{eqnarray}
H_{\text{min}} = \sum_i \big[ \frac{1}{2} J_{0}^{z} \mu^z_i + J_{0}^{\perp} \tau^x_i (\frac{1}{2} - \mu^z_i) \big] 
 + \sum_{\langle ij \rangle } \big[ J_1^z \tau^z_i \tau^z_j (1 + 4\mu_i^z \mu_j^z) 
 +2 J_2^z  \tau^z_i \tau^z_j  ( \mu_i^z + \mu_j^z) \big]  -\sum_i B\, \tau_i^z (1 + 2 \mu_i^z) . 
 \label{eqmin}
\end{eqnarray}
\end{widetext}
We would like to explain why we introduce the effective model first and then deal with the full model 
or the minimal model in terms of different spin variables. By directly examining 
the original full model in Eq.~\eqref{eq1} and/or the minimal model in Eq.~\eqref{eqmin}, 
one cannot extract much useful results. 
After the proper organization of the active Hilbert space and the definition/introduction of new effective 
variables, one could then make the connection with the effective quantum Ising model 
on the triangular lattice~\cite{PhysRevResearch.2.043013,Shen_2019,PhysRevB.109.L020403}. 
These procedures are the most insightful parts, and luckily the system is organized 
in such a remarkable way to naturally yield the effective model where the hidden physics can be revealed. 
The results of the 3-sublattice AF-singlet state, BKT physics, 3D XY quantum criticality, {\sl et al}
are all straightforward consequences from these procedures and observations.

With the useful results from $H_{\text{eff}}$, we analyze $H_{\text{min}}$ at zero field.
 Since $J_0^z$ is a very high energy scale and the ${\mu_i^z=1/2}$ states are high in energy,
 the ground state phase diagram is not altered. A bit more careful reasoning could solve 
$H_{\text{min}}$ by separating it into mutually-coupled $\mu$-channel model and $\tau$-channel 
model, and then obtain the ground states in this Kugel-Khomskii type of mean-field approaches~\cite{Supple}. 
The $\mu$-channel model gives the ground states with ${\mu_i^z =-1/2}$ on each site, and the
$\tau$-channel model is identical to $H_{\text{eff}}$ and thus has the same ground states. 
This treatment does not involve completely throwing away or projecting out the ${\mu_i^z=1/2}$ 
states and thus does not require a very strong $J_0^z$.

To obtain the excitations, we turn to the flavor-wave theory and make the detailed comparison with the linear spin-wave theory~\cite{Supple}.
The linear spin-wave theory only captures the separate spin flips of $\boldsymbol{\tau}$ and $\boldsymbol{\mu}$.
The linear flavor-wave theory has one extra mode per sublattice beyond the linear spin-wave theory 
by including the spin flipping process of for both $\boldsymbol{\tau}$ and $\boldsymbol{\mu}$.
For the 3-sublattice AF-singlet state, 9 flavor bosons are introduced. Keeping the quadratic terms
in the bosons for the minimal model, we obtain the linear flavor-wave model as
\begin{eqnarray}
&& H_{\text{fw}} = H_{\text{fw}\tau} + H_{\text{fw}\mu} + H_{\text{fw}\tau\mu}  ,\label{eq16} \\
&& H_{\text{fw}\tau} =
\sum_{ {\bf k} ,\mu\nu} \big[ 
               {\mathcal A}_{{\bf k} , \mu\nu}^{\phantom\dagger} a^\dagger_{{\bf k}\mu} a^{\phantom\dagger}_{ {\bf k} \nu} 
           +( {\mathcal B}_{{\bf k} ,\mu\nu}^{\phantom\dagger}   a^\dagger_{{\bf k}\mu} a^{\dagger}_{ {-{\bf k}}\nu}  + h.c.)
%  \nonumber \\  
 % && \quad\quad\quad\quad\quad  + \,  {\mathcal B}_{\mu\nu}^{\ast} ({\bf k}) a^{\phantom\dagger}_{{\bf k}\mu} a^{\phantom\dagger}_{{ -{\bf k}}\nu} 
\big], \,\,\,\,\,\,\,\, 
\nonumber
 \\
&& H_{\text{fw}\mu}=  
\sum_{ {\bf k}}  \big(
  \frac{J_0^z- s J_0^{\perp}  }{2}   \sum_{\mu \neq {\text B} }
 b^\dagger_{ {\bf k} \mu} b^{\phantom\dagger}_{{\bf k}\mu}  
 +  \frac{J_0^z +J_0^{\perp}}{2}  b^\dagger_{ {\bf k} {\text B}} b^{\phantom\dagger}_{{\bf k}{\text B}} 
\big), 
\nonumber 
\\
&& H_{\text{fw}\tau\mu}= \sum_{ {\bf k} } \big(   \frac{J_0^z-s J_0^{\perp} }{2}   +\frac{3}{4}  c^2 J_{\text{eff}}  \big) 
\sum_{\mu \neq {\text B} }
 c^\dagger_{ {\bf k} \mu} c^{\phantom\dagger}_{{\bf k}\mu}  \nonumber \\
 &&  \quad\quad\quad\quad\quad\quad\quad +  \sum_{ {\bf k} } \frac{J_0^z +J_0^{\perp}}{2}  c^\dagger_{ {\bf k} {\text B}} c^{\phantom\dagger}_{{\bf k}{\text B}}, \nonumber 
 \end{eqnarray}
where $\mu,\nu$ refer to the sublattices, ${\bf k}$ is summed over the magnetic Brillouin zone, $s=\sin \theta$ and $c=\cos \theta$
associate with the orientation of the magnetic order,
and we have the relation~\cite{Supple} 
${\mathcal A}_{{\bf k},\mu\nu} ={\mathcal A}^{\ast}_{{\bf k},\nu\mu}  , 
{\mathcal B}_{{\bf k},\mu\nu}   = {\mathcal B}^{\ast}_{{\bf k},\nu\mu} .
$

In $H_{\text{fw}}$, the spin dynamics within and out of the ${\mu^z=-1/2}$ manifold
is decoupled. $H_{\text{fw}\tau}$ describes the $\tau$ dynamics within the ${\mu^z=-1/2}$ 
manifold that is identical to the one for the effective model $H_{\text{eff}}$. 
 $H_{\text{fw}\mu}$ and $H_{\text{fw}\mu\tau}$ describe the spin fluctuations out of the ${\mu^z=-1/2}$ manifold. 
The full excitations of the 3-sublattice AF-singlet state are plotted in Fig.~\ref{fig3}(b), 
as well as the ones at the critical point in Fig.~\ref{fig3}(c). In Fig.~\ref{fig3}(b), the presence of the $\langle \tau^z \rangle$ order split
the excitations out of the ${\mu^z=-1/2}$ manifold~\cite{Supple}. 
With the similar and simpler approaches, we further established the full excitations 
for the dimer singlet in Fig.~\ref{fig3}(d), and
the excitations within and out of the ${\mu^z=-1/2}$ manifold become separately degenerate.

\emph{Discussion.}---We have obtained the key properties 
of the spin model for a dimer triangular lattice antiferromagnet by properly organizing 
the local spin Hilbert space and defining new sets of pseudospin variables. 
The hidden physics is then revealed 
with these variables. The full model can be thought as the quantum 
Ising model of the pseudospin-$\boldsymbol{\tau}$ decorated with the pseudospin-$\boldsymbol{\mu}$, 
and is solved with the mean-field approach and supplied with the existing understanding. 
This kind of decorated spin models with two distinct spins is potentially related to the 
decorated domain/string realizations of symmetry-protected topological orders~\cite{Savary2021,Chen_2014}.
In this work, we have mainly dealt with the zero-field case, the characteristic quantum 
excitations of each quantum phase are established. 
Although the transformation into the new sets of spin operators makes a great simplification
of the problem, the appearance of the original spin operators
becomes rather complex and behaves differently for different spin variables. 
This may cause some complication in the interpretation of the spectroscopic measurement. 
For both the 3-sublattice AF-singlet state and the dimer singlet, the excitations within and out of  
the ${\mu^z=-1/2}$ manifold are well separated from the separation of $a,b,c$ bosons in 
Eq.~\eqref{eq16}. From the relations of the physical spins and the flavor-wave bosons, 
the excitations of the $a,b,c$ bosons can be detected with inelastic neutron scattering
as the well-defined flavor-wave modes. For the 3-sublattice AF-singlet state, a subtle feature 
should occur. As the system orders in $\tau^z$, this longitudinal component correlation naturally
involves the 2-boson continuum of the $a$-bosons~\cite{Supple}.

% the excitation This naturally involves more than one flavors of Holstein-Primarkoff bosons, andthe two-magnon continuum can be induced in the spin correlation. 

Due to the strong Ising anisotropy, the system is expected to support the magnetization plateau. 
This is realized when the Zeeman energy gain overcomes the dimer singlet formation and the 
subdominant exchange energy remains optimized or nearly optimized. 
Unlike the quantum Ising model on the triangular lattice where the magnetic field simply
lifts the degeneracy of the up-up-down and down-down-up spin configurations on the triangular plaquettes
and does not alter the Ising exchange, 
here due to the extra $\mu$-spin degrees of freedom and the bilayer triangular geometry, the polarization
of the spin dimer from the singlet to the triplet actually ``quenches'' the neighboring exchange bonds, 
generating the emergent lattices. The resulting plateau states are briefly discussed~\cite{Supple}. 
Finally, since our minimal model does not have the fermion sign problem 
for the quantum Monte Carlo simulation, we expect more quantitative information 
to be established in later numerical simulation.

\emph{Acknowledgments.}---We would like to thank Tong Chen for very recent communication. 
This work is supported by the National Science Foundation of China with Grant No.~92065203 
and by the Ministry of Science and Technology of China with Grants No.~2021YFA1400300, 
and by the Fundamental Research Funds for the Central Universities, Peking University. 

\bibliography{refs}

\newpage

\end{document}

\title{Supplementary materials for ``Emergent Berezinskii-Kosterlitz-Thouless and Kugel-Khomskii physics in the triangular lattice bilayer colbaltate''}

 \author{Gang V. Chen}
 \thanks{Away from University of Hong Kong}
\affiliation{International Center for Quantum Materials, School of Physics, Peking University, Beijing 100871, China}
\affiliation{Collaborative Innovation Center of Quantum Matter, 100871, Beijing, China}
 
\begin{abstract}
\end{abstract}
\maketitle

\section{Mean-field theory for the effective model}

This section is devoted to explaining the mean-field treatment for the effective model $H_{\text{eff}}$ in the main text. 
For the three-sublattice AF-singlet state, we introduce the mean-field order parameters in Tab.~\ref{tab2}. 
In fact, this assignment of the order parameters captures both the three-sublattice AF-singlet state
and the dimer singlet (or the Ising disordered state). When $\theta = -\pi/2$, the system is in the dimer singlet. 

\begin{table}[h]
\begin{tabular}{ccc}
\hline\hline
& $\langle \tau^{z}_i \rangle $ & $\langle \tau^{x}_i \rangle $ 
\vspace{1.5mm}
\\
$i \in$ A sublattice & \quad $ \frac{1}{2} \cos \theta  $\quad & \quad $\frac{1}{2} \sin \theta $ \quad
\vspace{1.5mm}
\\
$i \in$ B sublattice & $0$ & $-\frac{1}{2}$
\vspace{1.5mm}
\\
$i \in$ C sublattice & $ -\frac{1}{2} \cos \theta $ & $\frac{1}{2} \sin \theta $ 
\vspace{1.5mm}
\\
\hline\hline
\end{tabular}
\caption{The mean-field order parameters for each sublattices in the 3-sublattice AF-singlet phase. Here $\theta <0 $. } 
\label{tab2}
\end{table}

In the Weiss mean-field treatment, $H_{\text{eff}}$ is reduced to 
\begin{eqnarray}
H_{\text{MF}} &=& \sum_i J_{0}^{\perp} \tau^x_i + \sum_{\langle ij \rangle} J_{\text{eff}}  \, \tau^z_i  \langle  \tau^z_j  \rangle \\
&=& \sum_{i \in \text{A}}  \big[ J_{0}^{\perp} \tau^x_i  - \frac{3}{2} J_{\text{eff}} \cos \theta \, \tau^z_i \big]   
+ \sum_{i \in \text{B}} J_{0}^{\perp} \tau^x_i \nonumber \\
&+&  \sum_{i \in \text{C}}  \big[ J_{0}^{\perp} \tau^x_i  + \frac{3}{2} J_{\text{eff}} \cos \theta  \, \tau^z_i \big] ,
\end{eqnarray}
where all the three sublattices are effectively decoupled. The self-consistent mean-field equations
for the A and C sublattices are the same. The self-consistent mean-field equation for the B sublattice 
is automatically solved as $\tau^x_i=-1/2$. 
The self-consistent mean-field equation for $i\in$ A is then given as
\begin{eqnarray}
\langle \tau^z_i \rangle  = \frac{\frac{3}{4} J_{\text{eff}} \cos \theta }{\big[(\frac{3}{2} J_{\text{eff}} \cos \theta )^2 +(J_{0}^{\perp} )^2 \big]^{1/2}},
\end{eqnarray}
 which yields the critical value of $J_0^{\perp}/ J_{\text{eff}} $ as 1.5. The numerical value 
 was found to be $\sim 0.82$~\cite{PhysRevB.68.104409,PhysRevResearch.2.043013}. 
 To capture the essential and qualitative physics and be consistent throughout, 
 we stick to the mean-field calculation and explain the quantitative aspect afterwards based on 
 the physical understanding. 
 
 % finite temperature transition 
 
 % BKT transition (check literature) 
 
\section{Linear spin-wave theory for the effective model}
\label{sec2}

In this section, we perform the linear spin-wave theory to solve for the excitations of the effective model $H_{\text{eff}}$
in the 3-sublattice AF-singlet state and the dimer singlet. 

\subsection{Three-sublattice AF-singlet state}
\label{secB1}

\begin{figure}
\includegraphics[width=7.5cm]{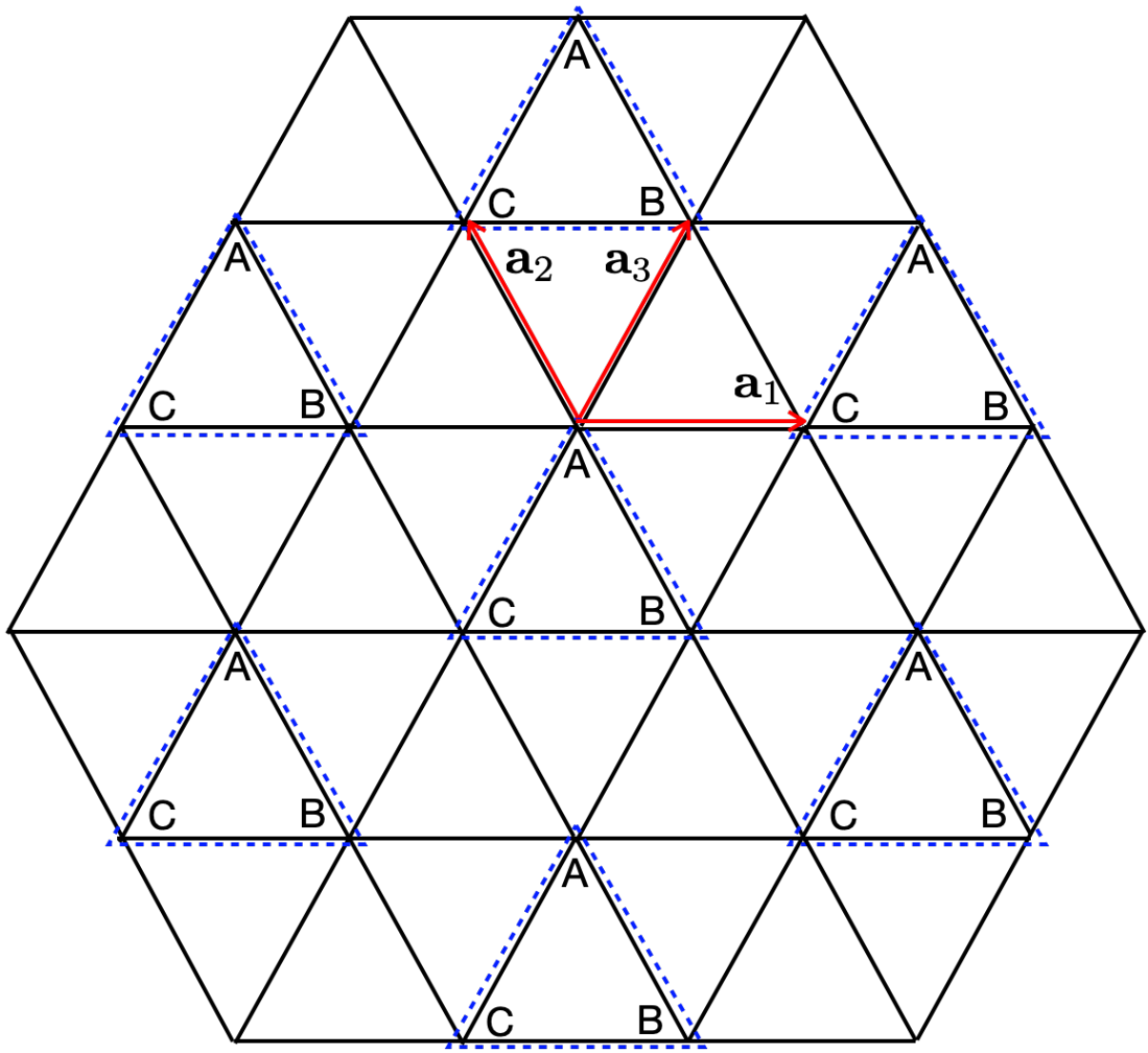}
\caption{The specification of magnetic sublattices and magnetic unit cells. The dashed triangle specifies the magnetic unit cell such that the 
intra- and inter- magnetic unit cell couplings become clear. 
The vectors ${\bf a}_1,{\bf a}_2, {\bf a}_3$ are three nearest-neighbor vectors 
of the triangular lattice.}
\label{sublattice}
\end{figure}

For both A and C sublattices, the pseudospin vectors point away from $x$ and $z$ directions, and we need to choose the 
quantization axis to be aligned with the ordered moment in the Holstein-Primarkoff spin wave representation. In the linear 
spin-wave formulation, the pseudospin operators are represented as 
\begin{eqnarray}
\left\{ 
\begin{array}{l}
\tau^x_i = + \frac{1}{2} \cos \theta (a^{\dagger}_{i\text{A}} +a^{}_{i\text{A}}) + \sin \theta (\frac{1}{2} -a^{\dagger}_{i\text{A}} a^{}_{i\text{A}} ) ,
\vspace{2mm}
\\
\tau^z_i = -\frac{1}{2} \sin \theta (a^{\dagger}_{i\text{A}} +a^{}_{i\text{A}}) + \cos \theta (\frac{1}{2} -a^{\dagger}_{i\text{A}} a^{}_{i\text{A}} ) ,
\end{array}
\right.
\end{eqnarray}
for $i \in $ A sublattice, 
\begin{eqnarray}
 \left\{ 
\begin{array}{l}
\tau^x_i = -  \frac{1}{2} + a^{\dagger}_{i\text{B}} a^{}_{i\text{B}} ,
\vspace{2mm}
\\
\tau^z_i = - \frac{1}{2} ( a^{\dagger}_{i\text{B}}+  a^{}_{i\text{B}}) ,
\end{array}
\right.
\end{eqnarray}
for $i \in $ B sublattice, 
and
\begin{eqnarray}
\left\{ 
\begin{array}{l}
\tau^x_i = - \frac{1}{2} \cos \theta (a^{\dagger}_{i\text{C}} +a^{}_{i\text{C}}) + \sin \theta (\frac{1}{2} -a^{\dagger}_{i\text{C}} a^{}_{i\text{C}} ) ,
\vspace{2mm}
\\
\tau^z_i = -\frac{1}{2} \sin \theta (a^{\dagger}_{i\text{C}} +a^{}_{i\text{C}}) - \cos \theta (\frac{1}{2} -a^{\dagger}_{i\text{C}} a^{}_{i\text{C}} ) ,
\end{array}
\right.
\end{eqnarray}
for $i \in $ C sublattice. According to the specification in Fig.~\ref{sublattice}, there are three intra-magnetic cell couplings and 
twelve inter-magnetic cell couplings. Keeping the quadratic terms in the bosons and properly grouping these terms, we obtain the 
linear spin wave theory,
\begin{eqnarray}
H_{\text{eff}} &\rightarrow& \sum_{\bf k} \sum_{\mu\nu} 
 a^{\dagger}_{{\bf k} \mu} a^{\phantom\dagger}_{{\bf k} \nu} {\mathcal A}_{\mu\nu} ({\bf k})
 + a^{\dagger}_{{\bf k}\mu} a^{\dagger}_{-{\bf k} \nu} {\mathcal B}_{\mu\nu} ({\bf k}) 
 \nonumber 
 \\
 && \quad\quad\quad\quad + \, a^{\dagger}_{{\bf k}\mu} a^{\dagger}_{-{\bf k} \nu} {\mathcal B}^{\ast}_{\mu\nu} ({\bf k}) 
\end{eqnarray}
where $\mu, \nu= \text{A, B, C}$, and the boson hoppings and pairings are given as

\begin{eqnarray}
{\mathcal A}_{11} ({\bf k}) &=& - J_0^{\perp} \sin \theta + \frac{3}{2} J_{\text{eff}} \cos^2 \theta  ,\\
{\mathcal A}_{22} ({\bf k}) &=& J_0^{\perp} , \\
{\mathcal A}_{33} ({\bf k}) &=& - J_0^{\perp} \sin \theta + \frac{3}{2} J_{\text{eff}} \cos^2 \theta \\
{\mathcal A}_{12}  ({\bf k}) &=& \frac{J_{\text{eff}}}{4}  \sin \theta ( e^{-i {\bf k}\cdot {\bf a}_1} + e^{-i {\bf k}\cdot {\bf a}_2} + e^{i {\bf k}\cdot {\bf a}_3}), \\
{\mathcal A}_{13} ({\bf k}) &=& \frac{J_{\text{eff}}}{4}  \sin^2 \theta ( e^{i {\bf k}\cdot {\bf a}_1} + e^{i {\bf k}\cdot {\bf a}_2} + e^{-i {\bf k}\cdot {\bf a}_3}) ,
\\
{\mathcal A}_{23} ({\bf k}) &=&   \frac{J_{\text{eff}}}{4}   \sin \theta ( e^{-i {\bf k}\cdot {\bf a}_1} + e^{-i {\bf k}\cdot {\bf a}_2} + e^{i {\bf k}\cdot {\bf a}_3}) ,
\end{eqnarray}
and
\begin{eqnarray}
{\mathcal B}_{12}({\bf k}) &=&  \frac{J_{\text{eff}}}{4}   \sin \theta  ( e^{-i {\bf k}\cdot {\bf a}_1} + e^{-i {\bf k}\cdot {\bf a}_2} + e^{i {\bf k}\cdot {\bf a}_3}) ,\\
{\mathcal B}_{13}({\bf k}) &=&  \frac{J_{\text{eff}}}{4}   \sin^2 \theta ( e^{i {\bf k}\cdot {\bf a}_1} + e^{i {\bf k}\cdot {\bf a}_2} + e^{-i {\bf k}\cdot {\bf a}_3}),
\\
{\mathcal B}_{23}({\bf k}) &=&  \frac{J_{\text{eff}}}{4}  \sin \theta  ( e^{-i {\bf k}\cdot {\bf a}_1} + e^{-i {\bf k}\cdot {\bf a}_2} + e^{i {\bf k}\cdot {\bf a}_3}) .
\end{eqnarray}

\subsection{Dimer singlet state}

For the dimer singlet, there is only one sublattice, 
and the spin-wave representation is much simpler and is given as 
\begin{eqnarray}
\left\{
\begin{array}{ll}
\tau^x_i = -\frac{1}{2} + a_i^\dagger a_i^{},
\vspace{2mm}
 \\
\tau^z_i = -\frac{1}{2} {(a^\dagger_i + a_i^{})}, 
\end{array}
\right.
\label{eqb14}
\end{eqnarray}
and the linear spin-wave Hamiltonian is then obtained as 
\begin{eqnarray}
H_{\text{eff}}& \rightarrow  & \frac{J_{\text{eff}}}{4} \sum_{\bf k  } 
\sum_{\mu } \cos  ( {\bf k} \cdot {\bf a}_{\mu}) ( a^{\phantom\dagger}_{\bf k} a^{\phantom\dagger}_{-{\bf k}} 
+a^{\dagger}_{\bf k} a^{\dagger}_{-{\bf k}}  + 2 a^\dagger_{\bf k} a_{\bf k}^{\phantom\dagger})
\nonumber \\
&& + J_0^{\perp}   \sum_{\bf k } 
a^\dagger_{\bf k} a_{\bf k}^{\phantom\dagger} ,
\end{eqnarray}
where the momentum summation is over the Brillouin zone of the triangular lattice. 
The excitation spectrum is ready to be solved and is given as
\begin{eqnarray}
\epsilon_{\bf k } &=& 2  \Big[ \big( \sum_{\mu} \frac{1}{4} J_{\text{eff}} \cos({\bf k}\cdot {\bf a}_{\mu}) + \frac{1}{2} J_0^{\perp}
 \big)^2 
\nonumber \\ 
&& \quad\quad\quad - \big(\sum_{\mu} \frac{1}{4} J_{\text{eff}} \cos({\bf k}\cdot {\bf a}_{\mu}) \big)^2
 \Big]^{1/2} . 
\end{eqnarray}
The dispersion arises from the effective exchange $J_{\text{eff}}$, and becomes flat 
when $J_{\text{eff}} \rightarrow 0$.

%\section{Finite-temperature mean-field theory}
 
 % discusss transition temperature 
 % vortex binding .

\section{Mean-field theory for the minimal model} 

We extend the mean-field treatment for the effective model $H_{\text{eff}}$ to 
the minimal model $H_{\text{min}}$. This mean-field treatment requires extra steps. 
One can actually adopt some experiences from the Kugel-Khomskii spin-orbital exchange model
where one can decouple the full model into the mutually-dependent spin and orbital mean-field models.
Here, we decouple the minimal model $H_{\text{min}}$ with coupled pseudospin $\tau_i$ and $\mu_j$ 
into the mutually-dependent mean-field models, and the models are given as 
\begin{eqnarray}
H_{\mu\text{MF}}  &=& \sum_i \frac{1}{2} J_0^z \mu_i^z + J_0^{\perp} \langle \tau_i^x \rangle (\frac{1}{2} - \mu_i^z) 
+ \sum_{\langle ij \rangle} \langle \tau_i^z \tau_j^z \rangle 
\nonumber \\
&&\times   \big[ J_1^z (1 + 4 \mu_i^z \mu_j^z) +2J_2^z  
(\mu_i^z + \mu_j^z ) \big] ,\\
H_{\tau\text{MF}} &=& \sum_i   J_0^{\perp}  \tau_i^x  [\frac{1}{2} - \langle  \mu_i^z \rangle ] +\sum_{\langle ij \rangle} 
 \tau_i^z \tau_j^z
\nonumber \\
&& \times   \big[  J_1^z (1   +  4   \langle \mu_i^z \mu_j^z \rangle )+ 2 J_2^z ( \langle \mu_i^z \rangle + \langle \mu_j^z \rangle) \big] . 
\end{eqnarray}
Since $J_0^z$ is the dominant energy scale, $H_{\mu\text{MF}}$ can be viewed as 
the Ising model with non-uniform Ising couplings in a strong ``magnetic field''. 
The Ising coupling depends on the bond correlation $\langle\tau_i^z \tau_j^z  \rangle$. 
$H_{\tau\text{MF}}$ should be viewed as a transverse field Ising model with the bond-dependent 
exchange couplings and site-dependent local transverse field. 

Here we explain that $\mu_i^z$ should be polarized to $-1/2$. First of all, 
a strong ``magnetic field'' of $J_0^z$ should already be sufficient to polarize the pseudospin $\mu_i^z$ as $\mu_i^z = -1/2$. 
Nevertheless, one can further soften the condition. From $H_{\tau\text{MF}}$, the effective transverse field on $\tau^x$ is 
always positive. This is due to the fact that
\begin{eqnarray}
\frac{1}{2} - \langle  \mu_i^z \rangle >0.
\end{eqnarray}
 The extreme case with $ {\mu_i^z=1/2}$ cannot 
happen as it is disfavored by a large $J_0^z$. 
Thus we have $\langle \tau^x_i \rangle<0$ throughout. 
In order to optimize the second term alone in $H_{\mu\text{MF}}$, we should require $\mu_i^z = -1/2$. 
The analysis of the third term in $H_{\mu\text{MF}}$ requires the knowledge of 
$\langle \tau_i^z \tau_j^z \rangle $ that 
can be considered from $H_{\tau\text{MF}}$. Although a strong $J_0^z$ would be sufficient to force
$\langle  \mu_i^z \rangle =-1/2$, since we are now softening this condition, a not-so-strong $J_0^z$ 
would simply favor a uniform $\langle  \mu_i^z \rangle <0$ with the support of the second term in $H_{\mu\text{MF}}$. 
This indicates that 
\begin{eqnarray}
\left\{
\begin{array}{l}
 J_1^z [ 1   +  4   \langle \mu_i^z \mu_j^z \rangle ] >J_1^z , 
 \vspace{2mm}  \\
  - 2 J_2^z  \leq 2 J_2^z [ \langle \mu_i^z \rangle + \langle \mu_j^z \rangle] <0 .
\end{array}
\right.
\end{eqnarray}
 From the exchange path and distances, it is expected that ${J_1^z \gg J_2^z}$. Thus,
 $H_{\tau\text{MF}}$ is the quantum Ising model with the antiferromagnetic 
 exchange, and this generally favors $\langle \tau_i^z \tau_j^z \rangle <0 $. 
 Thus, in the third term of $H_{\mu\text{MF}}$, the exchange coupling is ferromagnetic,
 and the overall coefficient of the $J_2$ term is ferromagnetic. Since $J_2$ is small, it cannot flip
 the sign of the total effective magnetic field for $\mu_i^z$. Thus, one can safely conclude that
 \begin{eqnarray}
 \mu_i^z = -\frac{1}{2}. 
 \label{eqmu}
 \end{eqnarray}
for all sites even if the dominant energy scale $J_0^z$ is reduced.  

With Eq.~\eqref{eqmu}, the mean-field model $H_{\tau\text{MF}}$ immediately becomes 
$H_{\text{eff}}$. Alternatively, one could simply set $ \mu_i^z = -\frac{1}{2}$ in the original 
minimal model $H_{\text{min}}$. The outcome is the same. But to obtain this outcome, one has 
to go through the above explanation. 

\section{Linear spin-wave theory for the minimal model}
\label{sec4}

Since the pseudspin ${\mu_i^z = -1/2}$ throughout the parameter regime of our interest, 
the Holstein-Primarkoff boson representation is the same in both 3-sublattice AF-singlet 
state and the Ising disordered state for the minimal model. We replace the pseudospin 
operator $\mu_i$ as 
\begin{eqnarray}
\mu^z_i =- \frac{1}{2} + b^\dagger_i b^{\phantom\dagger}_i, 
\end{eqnarray} 
and there is no other transverse component involved in $H_{\text{min}}$. For the full model
with finite $J_{1}^{\perp}$ and $J_2^{\perp}$, the transverse components are needed.

\subsection{Three sublattice order in the minimal model}

For the 3-sublattice AF-singlet state, the $\tau$-spins are represented in the same way as prescribed 
in Sec.~\ref{secB1}. As the transverse component of $\mu$ spins is not involved in $H_{\text{min}}$, no
dispersion is created. The resulting linear spin-wave model for the 3-sublattice AF-singlet is given as 
\begin{eqnarray}
H_{\text{sw}} = H_{\text{sw}\tau} + H_{\text{sw}\mu},
\end{eqnarray}
where $H_{\text{sw}\tau}$ is identical to the one in Sec.~\ref{secB1},
and $ H_{\text{sw}\mu}$ is given as
\begin{eqnarray}
 H_{\text{sw}\mu} &=& \sum_{\bf k}  \big[
  \frac{1}{2} ( J_0^z- J_0^{\perp} \sin \theta ) 
 (b^\dagger_{ {\bf k} {\text A}} b^{\phantom\dagger}_{{\bf k}{\text A}} + b^\dagger_{ {\bf k} {\text C}} b^{\phantom\dagger}_{{\bf k}{\text C}} )
 \nonumber \\
&& \quad\quad\quad\quad\quad  + \, \frac{1}{2} (J_0^z +J_0^{\perp}) b^\dagger_{ {\bf k} {\text B}} b^{\phantom\dagger}_{{\bf k}{\text B}} 
\big],
\end{eqnarray}
where the summation of the momentum is over the magnetic Brillouin zone. 
This creates three flat bands of $\mu$-sector excitations on top of the $\tau$-sector excitations. Among these three flat bands,
two are degenerate. 

\subsection{Dimer singlet state in the minimal model}

For the dimer singlet state, the $\tau$ component is represented in the same way as
the expression in Eq.~\eqref{eqb14}, and the minimal model is reduced to the spin wave model 
as
\begin{eqnarray}
H_{\text{min}} &\rightarrow & 
 \frac{J_{\text{eff}}}{4} \sum_{\bf k } 
\sum_{\mu } \cos  ({ {\bf k} \cdot {\bf a}_{\mu}}) ( a^{\phantom\dagger}_{\bf k} a^{\phantom\dagger}_{-{\bf k}} 
+a^{\dagger}_{\bf k} a^{\dagger}_{-{\bf k}}  + 2 a^\dagger_{\bf k} a_{\bf k}^{\phantom\dagger})
\nonumber \\
&+& \sum_{\bf k}  J_0^{\perp} a^\dagger_{\bf k} a^{\phantom\dagger}_{\bf k} + 
\sum_{\bf k} \frac{1}{2} (J_0^z + J_0^{\perp}) b^\dagger_{\bf k} b^{\phantom\dagger}_{\bf k} ,
\label{eq28}
\end{eqnarray}
where a flat excitation of the $\mu$ sector with the energy $\frac{1}{2} (J_0^z + J_0^{\perp})$ is introduced on top of the 
excitation of the effective model $H_{\text{eff}}$.

\section{Linear flavor-wave theory for the minimal model}

In this section, we implement the linear flavor-wave theory for the minimal model. 
The linear spin-wave theory for the minimal model captures the linear spin-flipping 
process of both $\boldsymbol{\tau}$ and $\boldsymbol{\mu}$ pseudospins separately, 
but does not capture the process of flipping $\boldsymbol{\tau}$ and $\boldsymbol{\mu}$
at the same time in the linear order. If one examines the spin operator such as 
$S^x_{ui}$ that is equal to $2\tau_i^x \mu^x_i$ in Tab. 1 of the main text, this spin operator 
would flip $\tau^z$ and $\mu^z$ simultaneously. To capture this part of excitation, 
a bit more complete treatment is to rely on the linear flavor-wave theory.

In fact, as we have viewed this system as one example of ``multiflavor Mott insulators'',
the key property of  ``multiflavor Mott insulators'' is that the system is more delocalized
on the local Hilbert space and allows the quantum tunneling from one local state to all 
other local states at the linear order of the Hamiltonian~\cite{Chen2021MultiflavorMI}.  
For such systems when they are ordered, the linear flavor-wave theory is more appropriate
than the linear spin-wave theory by capturing more relevant modes of excitations at the linear level. 
In the following, we first implement the linear flavor-wave theory for the 
dimer singlet and then solve the excitations for the 3-sublattice AF-singlet state.

\subsection{Linear flavor-wave theory for the dimer singlet} 

We describe the flavor-wave treatment for the dimer singlet here. 
A more complicated version can be straightforwardly obtained for the 3-sublattice AF-singlet state. 
The dimer singlet is a direct product state where every site can be 
described by the same wavefunction,
\begin{equation}
| {\text{dimer singlet}} \rangle \approx \prod_i | \tau_i^x =- \frac{1}{2}, \mu_i^z =- \frac{1}{2}  \rangle_i .
\end{equation}
Thus, we choose our basis for the flavor-wave bosons as 
\begin{eqnarray}
|0\rangle_i \equiv  a_{i0}^\dagger | \emptyset \rangle &\equiv & | \tau_i^x =- \frac{1}{2}, \mu_i^z =- \frac{1}{2}  \rangle , \\
|1\rangle_i \equiv a_{i1}^\dagger | \emptyset \rangle &\equiv & | \tau_i^x =+ \frac{1}{2}, \mu_i^z =- \frac{1}{2}  \rangle , \\
|2\rangle_i \equiv a_{i2}^\dagger | \emptyset \rangle &\equiv & | \tau_i^x =- \frac{1}{2}, \mu_i^z =+ \frac{1}{2}  \rangle , \\
|3\rangle_i \equiv a_{i3}^\dagger | \emptyset \rangle &\equiv & | \tau_i^x =+ \frac{1}{2}, \mu_i^z =+ \frac{1}{2}  \rangle , 
\end{eqnarray}
where $ | \emptyset \rangle$ is the vacuum state. A Hilbert space constraint on every site is imposed with
\begin{eqnarray}
a_{i0}^\dagger a^{}_{i0} + a_{i1}^\dagger a^{}_{i1} 
+ a_{i2}^\dagger a^{}_{i2} 
+a_{i3}^\dagger a^{}_{i3} =1 . 
\end{eqnarray}
The physical spin operators can be expressed as the bilinears of the flavor-wave bosons. For example, $S^x_{ui}$
is given as 
\begin{eqnarray}
S^x_{ui} \equiv  \sum_{m,n} {\tensor[^{}_i]{\langle m |}{} } S^x_{ui} | n \rangle_i \, a^\dagger_{im} a^{}_{in} ,
\label{eq35}
\end{eqnarray}
where $m,n = 0,1,2,3$. 

For the dimer singlet, the $a_{0}$ boson is condensed with 
\begin{eqnarray}
\langle a_{i0}^{\dagger} \rangle = \langle a_{i0}^{} \rangle &=& \sqrt{1 - a_{i1}^{\dagger} a_{i1}^{} - a_{i2}^{\dagger} a_{i2}^{}- a_{i3}^{\dagger} a_{i3}^{} } 
\nonumber \\
&\approx & 1 . 
\end{eqnarray}
The relevant spin operators in the minimal model can then be obtained as 
\begin{eqnarray}
&& \tau_i^x (\frac{1}{2} - \mu_i^z) \approx -\frac{1}{2} + a^\dagger_{i1} a^{}_{i1} + \frac{1}{2} ( a^\dagger_{i2} a^{}_{i2} + a^\dagger_{i3} a^{}_{i3}),
\\
&& \tau_i^z \mu_i^z \approx \frac{1}{4} ( a_{i1}^{} + a_{i1}^{\dagger} ) ,
\\
&& \tau_i^z \approx -\frac{1}{2} ( a_{i1}^{} + a_{i1}^{\dagger} ) ,
\\
&& \tau_i^x \approx -\frac{1}{2} + a^\dagger_{i1} a^{}_{i1}  + a^\dagger_{i3} a^{}_{i3} ,
\\
&& \mu_i^z \approx -\frac{1}{2} + a^\dagger_{i2} a^{}_{i2}  + a^\dagger_{i3} a^{}_{i3} .
\end{eqnarray}
One can make a direct comparison with the linear spin-wave theory in Sec.~\ref{sec2}
and Sec.~\ref{sec4}, and identify that 
\begin{eqnarray}
  a^{}_{i} = a_{i1}^{}, \quad\quad\quad
  b^{}_{i} = a_{i2}^{} .
\end{eqnarray}
Thus, to simplify the notation, we let $a_{i3}= c_i$ and continue to use the notations of $a_i$ and $b_i$. 

In this formulation of the linear flavor-wave theory, the minimal model is reduced to 
\begin{eqnarray}
H_{\text{min}} & \rightarrow & \frac{J_{\text{eff}}}{4} \sum_{\bf k } 
\sum_{\mu } \cos  ({ {\bf k} \cdot {\bf a}_{\mu}}) ( a^{\phantom\dagger}_{\bf k} a^{\phantom\dagger}_{-{\bf k}} 
+a^{\dagger}_{\bf k} a^{\dagger}_{-{\bf k}}  + 2 a^\dagger_{\bf k} a_{\bf k}^{\phantom\dagger})
\nonumber \\
&+& \sum_{\bf k}  J_0^{\perp} a^\dagger_{\bf k} a^{\phantom\dagger}_{\bf k} + 
\sum_{\bf k} \frac{1}{2} (J_0^z + J_0^{\perp}) 
(b^\dagger_{\bf k} b^{\phantom\dagger}_{\bf k} + c^\dagger_{\bf k} c^{\phantom\dagger}_{\bf k}). 
\end{eqnarray}
In comparison with the linear spin-wave theory in Eq.~\eqref{eq28}, 
the linear flavor-wave theory model has one extra $c$-boson 
whose energy is identical to the $b$-boson.

\subsection{Linear flavor-wave theory for the three-sublattice AF-singlet state} 

Here we turn to the linear flavor-wave theory for the 3-sublattice AF-singlet state. 
Here we describe the formulation for the A sublattice, 
and the C sublattice lattice is then obtained from the A sublattice 
by simply setting ${\theta \rightarrow \pi - \theta}$. 
For the B sublattice, since the local state is a dimer singlet, we simply use the 
formulation of the previous subsection.

For $i \in \,$A sublattice, we define the flavor-wave bosons as
\begin{eqnarray}
&&  | 0 \rangle_i \equiv  a_{i{\text A}0}^\dagger | \emptyset \rangle \equiv  | \tilde{\tau}_i^z =+\frac{1}{2},  \mu_i^z = -\frac{1}{2} \rangle ,
\\
&&  | 1 \rangle_i \equiv   a_{i{\text A}1}^\dagger | \emptyset \rangle \equiv  | \tilde{\tau}_i^z =-\frac{1}{2},  \mu_i^z = -\frac{1}{2} \rangle ,
\\
&&  | 2 \rangle_i \equiv  a_{i{\text A}2}^\dagger | \emptyset \rangle \equiv  | \tilde{\tau}_i^z =+\frac{1}{2},  \mu_i^z = +\frac{1}{2} \rangle ,
\\
&&  | 3 \rangle_i \equiv  a_{i{\text A}3}^\dagger | \emptyset \rangle \equiv  | \tilde{\tau}_i^z =-\frac{1}{2},  \mu_i^z = +\frac{1}{2} \rangle ,
\end{eqnarray}
and supplement with the Hilbert space constraint 
\begin{equation}
a_{i{\text A}0}^\dagger a^{}_{i{\text A}0} + a_{i{\text A}1}^\dagger a^{}_{i1} 
+ a_{i{\text A}2}^\dagger a^{}_{i{\text A}2} 
+a_{i{\text A}3}^\dagger a^{}_{i{\text A}3} =1 ,
\end{equation}
where $\tilde{\tau}^z_i \equiv \cos \theta \, \tau_i^z + \sin \theta \, \tau_i^x$, and 
\begin{eqnarray}
 | \tilde{\tau}_i^z =+\frac{1}{2} \rangle 
& =& \cos \frac{\theta}{2}  | \tau^z_i= \frac{1}{2} \rangle
  + \sin \frac{\theta}{2} | \tau^z_i =- \frac{1}{2}\rangle,  \\
   | \tilde{\tau}_i^z =-\frac{1}{2} \rangle 
& =& -\sin \frac{\theta}{2}  | \tau^z_i= \frac{1}{2} \rangle
  + \cos \frac{\theta}{2} | \tau^z_i =- \frac{1}{2}\rangle  .
\end{eqnarray}
For the A sublattice, $a_{i{\text A}0}$ is condensed, and we can approximately have
\begin{eqnarray}
\langle a_{i{\text A}0}^{} \rangle = \langle a_{i{\text A}0}^{\dagger} \rangle \approx 1,
\end{eqnarray}
and the relevant spin operators for the A sublattice are given as 
\begin{eqnarray}
&&\mu^z_i = -\frac{1}{2} + a^\dagger_{i{\text A}2} a^{}_{i{\text A}2} + a^\dagger_{i{\text A}3} a^{}_{i{\text A}3} , \\
&& \tau^x_i (\frac{1}{2} - \mu_i^z) = \frac{1}{2} \sin \theta  + \frac{1}{2} \cos \theta ( a^\dagger_{i{\text A}1} + a^{}_{i{\text A}1}) 
  - \sin \theta \,  a^\dagger_{i{\text A}1} a^{}_{i{\text A}1}  \nonumber \\
          && \quad\quad\quad\quad\quad\quad  - \frac{1}{2} \sin \theta (a^\dagger_{i{\text A}2} a^{}_{i{\text A}2} + a^\dagger_{i{\text A}3} a^{}_{i{\text A}3})  ,\\
               && \tau_i^z =  \frac{1}{2} \cos \theta - \cos \theta  (a^\dagger_{i{\text A}1} a^{}_{i{\text A}1} + a^\dagger_{i{\text A}3} a^{}_{i{\text A}3})  \nonumber \\
               &&  \quad\quad\quad\quad\quad\quad -\frac{1}{2} \sin \theta ( a^\dagger_{i{\text A}1} + a^{}_{i{\text A}1}  ) , \\
               && \tau^z_i \mu^z_i = -  \frac{1}{4} \cos \theta +   \frac{1}{2} \cos \theta  (a^\dagger_{i{\text A}1} a^{}_{i{\text A}1} + a^\dagger_{i{\text A}2} a^{}_{i{\text A}2})
               \nonumber \\
               &&  \quad\quad\quad\quad\quad\quad + \frac{1}{4} \sin \theta (a^\dagger_{i{\text A}1} + a^{}_{i{\text A}1} ).
\end{eqnarray}
A comparison with Sec.~\ref{sec2} and Sec.~\ref{sec4}, one can infer that 
\begin{eqnarray}
a_{i{\text A}} = a_{i{\text A}1}, \quad\quad\quad  b_{i{\text A}}=a_{i{\text A}2}. 
\end{eqnarray}
Likewise, for the B and C sublattices, we have 
\begin{eqnarray}
&& a_{i{\text B}} = a_{i{\text B}1}, \quad\quad\quad  b_{i{\text B}}=a_{i{\text B}2}, \\
&& a_{i{\text C}} = a_{i{\text C}1}, \quad\quad\quad  b_{i{\text C}}=a_{i{\text C}2}. 
\end{eqnarray}

With the above representation of the spin operators, we reduce the minimal model into the linear flavor-wave model as 
\begin{eqnarray}
H_{\text{fw}} = H_{\text{fw}\tau} + H_{\text{fw}\mu} + H_{\text{fw}\tau\mu}.
\end{eqnarray}
Since we have identified the relation between the flavor-wave boson and the spin-wave boson, we then can identify the 
relation on the model side and have
\begin{eqnarray}
&& H_{\text{fw}\tau}  = H_{\text{sw}\tau} , \\
&& H_{\text{fw}\mu}  = H_{\text{sw}\mu}  .
\end{eqnarray}
$H_{\text{fw}\tau\mu}$ is a new term in the linear flavor-wave theory that arises from the 
flipping of the both $\boldsymbol{\tau}$ and $\boldsymbol{\mu}$ pseudospins, and is given as 
\begin{eqnarray}
H_{\text{fw}\tau\mu} &=& \sum_{\bf k} ( \frac{J_0^z-\sin \theta J_{0}^{\perp}}{2} + \frac{3}{4} \cos^2 \theta J_{\text{eff}} ) \sum_{\mu \neq {\text B}} 
c^\dagger_{{\bf k} \mu} c^{}_{{\bf k}\mu} \nonumber \\
&& \quad\quad \quad + \sum_{\bf k} \frac{J_0^z + J_0^{\perp}}{2} c^\dagger_{{\bf k} {\text B}} c^{}_{{\bf k} {\text B}},
\label{eq62}
\end{eqnarray}
where we have replaced $a_{i{\text A}3}, a_{i{\text B}3}, a_{i{\text C}3} $ with $c_{i {\text A}}, c_{i {\text B}}, c_{i {\text C}}$, respectively. 
This result has been expressed in the main text. 
Although a simplified energy in the first line of Eq.~\eqref{eq62} can be obtained, we keep the expression that relates to the magnetic order
such that the connection to the dimer singlet state can be made. 
This $c$-boson excitation has the same energy as the $b$-boson in the dimer singlet state, but differs from the $b$-boson
in the 3-sublattice AF-singlet state for the A and C sublattices. This is because
the presence of the pseudospin $\boldsymbol{\tau}_i$ order (in the $z$ component) modulates the 
excitations from the ${\mu_i^z=-1/2}$ manifold to the ${\mu_i^z=1/2}$ manifold. Thus, we observe
 three flat energy bands in the excitations of the 3-sublattice AF-singlet state in Fig.~3 of the main text.

 \section{The spin components in the flavor-wave bosons}

In this section, we clarify the relations between the physical spins and the flavor-wave bosons in the linear flavor-wave approximation. 
These relations are particularly useful to understand the experimental signatures in the inelastic neutron scattering measurements. In the 
actual experiments, the experimental apparatus is directly coupled to the physical spins, instead of the effective pseudospins. This is our
motivation to establish these relations.

\subsection{Dimer singlet}

\begin{table}[h]
\begin{tabular}{p{1cm}p{1.5cm}p{2.1cm}}
\hline\hline 
$S^z_{ui}$ & $=\tau_i^z $ & $\approx -\frac{1}{2} ( a^{}_i + a^\dagger_i)$      \vspace{0.6mm} \\
$S^x_{ui}$ & $=2 \tau_i^x \mu_i^x $ & $\approx-\frac{1}{2} ( b^{}_i + b^\dagger_i)$   \vspace{0.6mm}\\
$S^y_{ui}$ & $=2 \tau_i^y \mu_i^x $ & $\approx+ \frac{i}{2} ( c^{}_i - c^\dagger_i)$  \vspace{0.6mm} \\
$S^z_{di}$ & $=2\tau^z_i \mu^z_i $ &  $\approx+ \frac{1}{2} ( a^{}_i + a^\dagger_i)$  \vspace{0.6mm}\\
$S^x_{di}$ & $=\mu^x_i $ &  $\approx+\frac{1}{2} ( b^{}_i + b^\dagger_i)$  \vspace{0.6mm}\\
$S^y_{di}$ & $=2\tau^z_i \mu^y_i $ & $\approx- \frac{i}{2} ( c^{}_i - c^\dagger_i)$  \vspace{0.6mm}  \\
\hline\hline
\end{tabular}
\caption{The representation of the physical spin operators in terms of flavor-wave bosons in the dimer singlet. 
The first column is the physical spin, the second column is in terms of the two pseudospins, 
and the third column is in the linear flavor-wave boson approximation. 
Here, we actually keep the flavor-wave bosons up to the quadratic order for the diagonal contribution. }
\label{tab2}
\end{table}

In Tab.~\ref{tab2}, we list the expression of the physical spin in the linear flavor-wave approximation. 
Following the expression in Eq.~\eqref{eq35}, we keep the off-diagonal part to the linear order in the flavor-wave boson
and keep the diagonal part to the quadratic order in the flavor-wave boson. One could improve the results by including the
high-order off-diagonal part. From Tab.~\ref{tab2}, one can immediately conclude that
all the $a,b,c,$ flavor bosons should occur as the well-defined flavor-wave modes in the inelastic neutron scattering measurements.

\subsection{Three-sublattice AF-singlet state}

For the 3-sublattice AF-singlet state, we list the results in Tab.~\ref{tab3} for the A sublattice, 
and the C sublattice is obtained by setting $\theta$ to $\pi- \theta$. 
For the B sublattice, we directly adopt the results in Tab.~\ref{tab2}. 

\begin{table}[h]
\begin{tabular}{p{0.5cm}p{7.9cm}}
\hline\hline 
$S^z_{ui}$ &  $\approx \frac{1}{2} \cos \theta   -\frac{1}{2}\sin \theta ( a^{}_{i{\text A}} + a^\dagger_{i{\text A}}) - \cos \theta (a^\dagger_{i{\text A}} a_{i{\text A}}^{} + c^\dagger_{i{\text A}} c_{i{\text A}}^{}  )$ \vspace{1mm} \\
$S^x_{ui}$ &   $\approx \frac{1}{2} \sin \theta ( b^{}_{i{\text A}}  + b^\dagger_{i{\text A}} ) + \frac{1}{2} \cos \theta ( c^{}_{i{\text A}}  + c^\dagger_{i{\text A}} )$  \vspace{1mm} \\
$S^y_{ui}$ &   $\approx- \frac{i}{2} ( c^{}_{i{\text A}} - c^\dagger_{i{\text A}})$  \vspace{1mm} \ \\
$S^z_{di}$ &   $\approx -\frac{1}{2} \cos \theta   +\frac{1}{2}\sin \theta ( a^{}_{i{\text A}} + a^\dagger_{i{\text A}}) + \cos \theta (a^\dagger_{i{\text A}} a_{i{\text A}}^{} + b^\dagger_{i{\text A}} b_{i{\text A}}^{}  )$ \vspace{1mm} \\
$S^x_{di}$ &    $\approx+\frac{1}{2} ( b^{}_{i{\text A}} + b^\dagger_{i{\text A}})$\vspace{1mm}\\
$S^y_{di}$ &  $\approx +  \frac{i}{2} \cos \theta (b_{i{\text A}}^{} - b_{i{\text A}}^{\dagger}) - \frac{i}{2} \sin \theta ( c^{}_{i{\text A}} - c^\dagger_{i{\text A}})$ \vspace{1mm} \\
\hline\hline
\end{tabular}
\caption{The representation of the physical spin operators in terms of flavor-wave bosons 
in the 3-sublattice AF-singlet state for the A sublattice. 
The first column is the physical spin, and the second column is in the linear flavor-wave boson approximation. }
\label{tab3}
\end{table}

Based on these relations, all the $a,b,c,$ flavor bosons should occur as the well-defined flavor-wave modes in the inelastic neutron scattering measurements. 
Moreover, the $z$-component spin correlation will involve the $a$-boson continuum in the inelastic neutron scattering measurements. The reason is as follows.
To diagonalize $H_{\text{fw}\tau}$, one needs to make a Bogoliubov transformation due to the pairing term. Thus, the action of $a^\dagger_i a_i$ on the 
ground state has the zero-point contribution and is no-zero, and the $z$-component spin correlation then contains the $a$-boson continuum. 
For the $b$ or $c$ bosons, the $H_{\text{fw}\mu}$ and $H_{\text{fw}\mu\tau}$ are diagonalized without any transformation. The action of $b^\dagger_i b_i$
(or $c^\dagger_i c_i$) on the ground state is zero.

\begin{figure}[t]
\includegraphics[width=7cm]{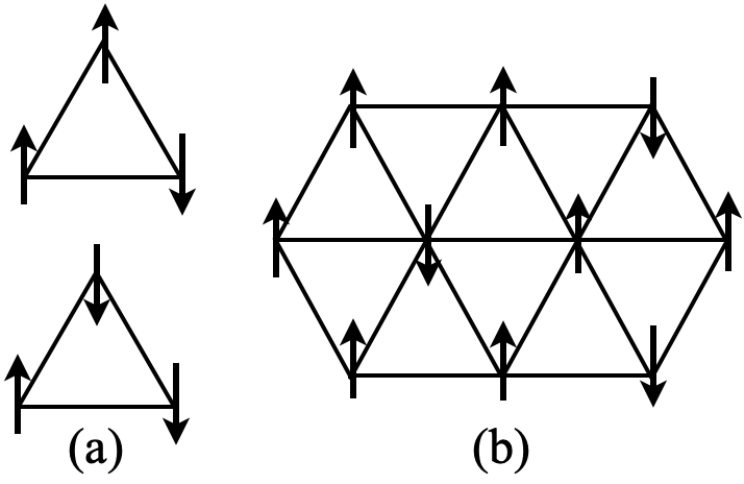}
\caption{(a) The degenerate up-up-down and down-down-up spin configurations 
      for the classical Ising model on the triangular plaquette. 
(b) The 1/3 magnetization plateau state on the triangular lattice. }
\label{sfig1}
\end{figure} 

\section{Magnetization plateau states}

Here we give a discussion of the magnetization plateau states in the presence of $z$-direction magnetic fields. 
To motivate this discussion, we first sketch the 1/3 magnetization plateau state for the antiferromagnetic Ising model on 
the triangular lattice under the external magnetic field
on the Ising component. The Zeeman coupling over there immediately lifts the degeneracy of the
``down-down-up'' and ``up-up-down'' spin configurations for each triangular plaquette (see Fig.~\ref{sfig1}(a)).
Once the ``up-up-down'' spin configuration is fixed on one triangular plaquette, the spin configuration
on the whole triangular lattice is determined (see Fig.~\ref{sfig1}(b)), and the exchange energy 
on each triangular plaquette is optimized. Such a ``up-up-down'' state retains the three-sublattice 
structure and gives rise to the 1/3 magnetization plateau on the triangular lattice. 
Due to the robustness of this ``up-up-down'' state, it persists even to the quantum case 
with the transverse field~\cite{PhysRevResearch.2.043013}.

\begin{figure}[t]
\includegraphics[width=5.6cm]{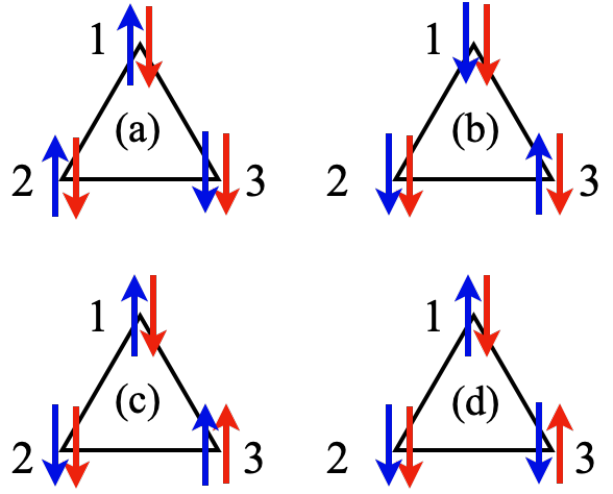}
\caption{
(a) and (b) are the $\tau^z$ Ising spin configurations with $\mu_i^z= -1/2$ throughout on one triangular plaquette. 
(c) and (d) are the $\tau^z$ Ising spin configurations by flipping one $\mu_i^z$ from $-1/2$ to $+1/2$.
The blue arrows are for $\tau^z$ spins, and the red arrows are for $\mu^z$ spins. The same choice is given for the other figures.  }
\label{fig6}
\end{figure}

\begin{figure}[t]
\includegraphics[width=8.4cm]{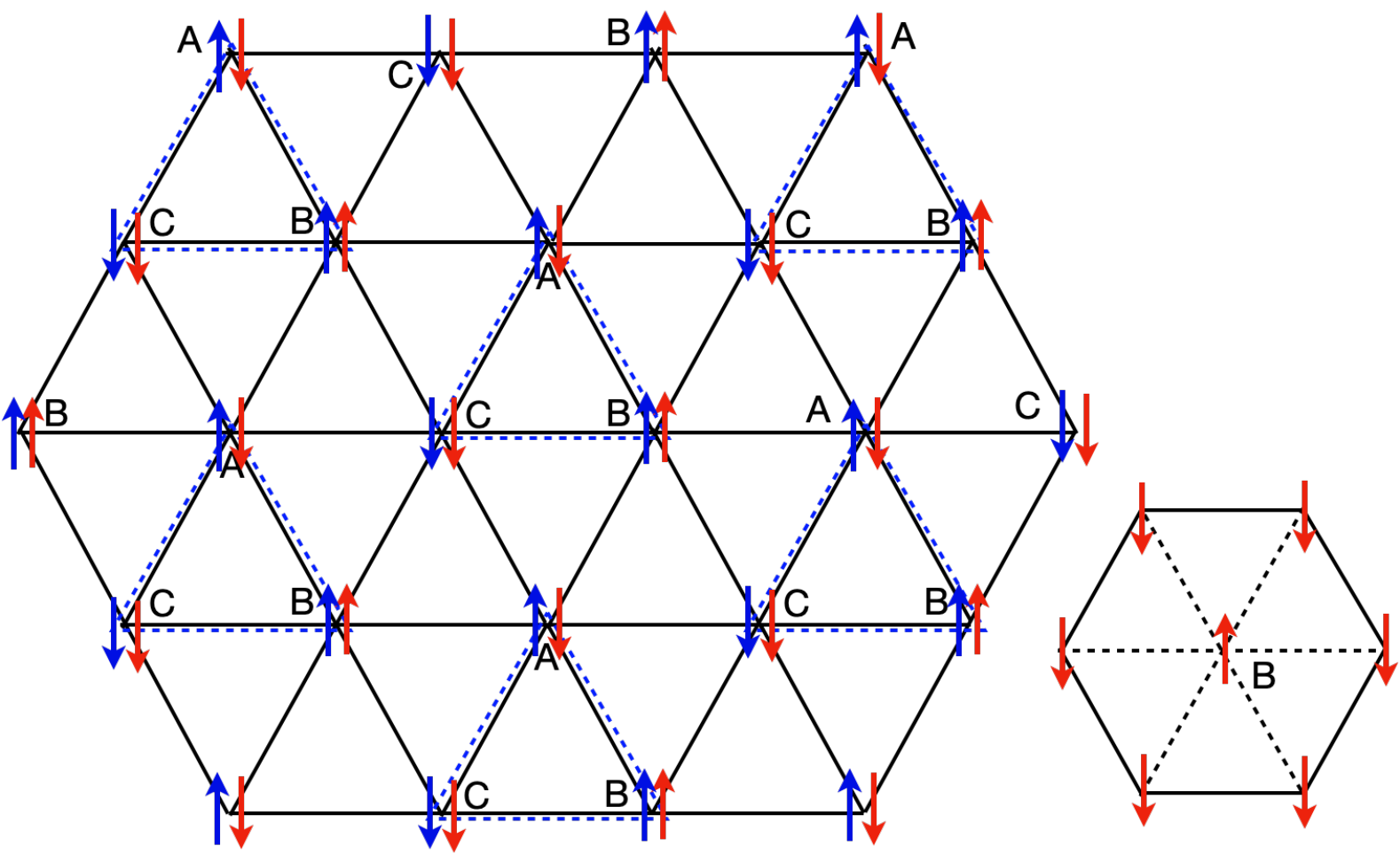}
\caption{The spin configuration 
for the 1/3 magnetization plateau of the minimal model by setting $J_0^{\perp} =0$. 
The B sublattice has $\mu_i^z = 1/2$. 
In the inset, the dashed bonds have zero exchange coupling for $\tau^z$'s, and the remaining
hexagon with {\sl even} number of bonds is antiferromagnetically coupled for the $\tau^z$ spins. 
}
\label{fig7}
\end{figure}

For the minimal model $H_{\text{min}}$, the local Hilbert space is large, 
and this is where the complication and difference arise.   
We consider the classical limit by setting ${J_{0}^{\perp} =0}$, 
and the regime with $J_0^z \gg J_1^z  \gg J_2^z$. At the zero magnetic field, 
$\mu^z=-1/2$ on every site, and the model is the antiferromagnetic Ising model with $\tau$ variables. 
For each triangular plaquette, there exist one unsatisfied bond and two satisfied bonds, 
and overall there is one net satisfied bonds for the Ising spin interactions of the $\tau$ 
spins. This is shown as Figs.~\ref{fig6}(a) and (b). 
When the external magnetic field polarizes one dimer singlet into the dimer triplet 
by flipping one $\mu_i^z$ from $-1/2$ to $+1/2$, the exchange coupling on the antiferromagnetically
arranged $\mu^z$ bond is quenched. 
In Figs.~\ref{fig6}(c) and (d), the exchange interactions on the 13 and 23 bonds are zero,
and the remaining $23$ bond is optimized. Thus, the net satisfied bond on the triangular plaquette
remains to be one. The spin state in Fig.~\ref{fig6}(d) does not gain Zeeman energy for the positive-$\hat{z}$
magnetic field compared to the
spin state in Fig.~\ref{fig6}(c) because the total magnetization on the site $3$ is down. 
 Once the Zeeman energy gain of the spin state in Fig.~\ref{fig6}(c) 
overcomes the singlet-triplet energy gap $J_0^z/2$, then the spin state is satisfied. 
To make sure only one $\mu_i^z$ spin to be flipped for each triangular plaquette 
for the whole triangular lattice, one arrange the $\mu^z=1/2$ sites in a next-nearest neighbor
fashion, which is very much like the ``up'' spin of the 1/3 magnetization state in Fig.~\ref{sfig1}(b). 
As we show in Fig.~\ref{fig7}, the B sites have $\mu^z=1/2, \tau^z_i=1/2$ to gain the Zeeman energy, and are decoupled from the remaining 
sites due to the quenched exchange interactions. The remaining sites have $\mu^z=-1/2$
and form an emergent honeycomb lattice that are unfrustrated (see the inset of Fig.~\ref{fig7}).  
These sites have an antiferromagnetically
arranged $\tau^z$ spin configuration, and there is no extensive or subextensive degeneracy any more. 
The magnetization of this state is solely contributed from the B sites
and is equal to 1/3 of the fully polarized state. Due to the robustness of this state,
the system should have the 1/3 magnetization plateau for this state. 
As it is depicted in Fig.~\ref{fig7}, this state retains the three-sublattice 
structure. If one further increases the magnetic field, more plateau states will be generated.

\bibliography{refs}